\documentclass[a4paper,11pt]{article}
\pdfoutput=1

\usepackage{mathtools}
\usepackage{cases}
\usepackage{booktabs}
\usepackage{jcappub}

\usepackage{mathrsfs}

\usepackage{MnSymbol}

\usepackage[T1,T2A]{fontenc}
\usepackage[utf8]{inputenc}
\usepackage[russian,english]{babel}

\usepackage{newunicodechar}

\newunicodechar{℈}{\azeriE}
\newunicodechar{ə}{\azerie}

\DeclareRobustCommand{\azeriE}{%
  {\fontencoding{T2A}\selectfont\symbol{"9A}}%
}
\DeclareRobustCommand{\azerie}{%
  {\fontencoding{T2A}\selectfont\symbol{"BA}}%
}

\makeatletter
\expandafter\def\expandafter\@uclclist\expandafter
  {\@uclclist\azerie\azeriE}
\makeatother

\newcommand{\rar}{\rightarrow}

\newcommand{\pd}{\partial}
\newcommand{\dd}{\mathrm{d}}

\newcommand{\tr}{t_r}
\newcommand{\trt}{t_r(t)}
\newcommand{\trHel}{\left<t^2_r(t)\right>_\text{hel}}
\newcommand{\trSq}{\sqrt{\left<t^2_r(t)\right>}}

\newcommand{\al}{\alpha}

\newcommand{\de}{\delta}

\newcommand{\rh}{\rho}
\newcommand{\si}{\sigma}

\newcommand{\vep}{\varepsilon}

\newcommand{\vth}{\vartheta}
\newcommand{\vph}{\varphi}

\newcommand{\vn}{\mathbf{n}}
\newcommand{\vx}{\mathbf{x}}

\newcommand{\vepT}{\vep_{\rm T}}
\newcommand{\vepV}{\vep_{\rm V}}
\newcommand{\vepS}{\vep_{\rm S}}

\newcommand{\rhoDM}{\rho_{\text{DM}}}

\newcommand{\deq}{\coloneqq}

\newcommand{\Eq}[1]{eq.~(\ref{#1})}

\newcommand{\nn}{\nonumber} 

\newcommand{\M}{M}
\newcommand{\Mij}{\M_{ij}}

\newcommand{\Mmn}{\M_{\mu\nu}}
 
\newcommand{\mpl}{M_{\text{P}}}
\newcommand{\D}{\nabla}
\newcommand{\ldB}{\lambda_\text{dB}}

\usepackage[small,labelfont=bf,textfont=it]{caption}
\usepackage{float}

\usepackage{newfloat}
\DeclareFloatingEnvironment[
    fileext=lop,
    listname={List of Panels},
    name=Panel,
    placement=tbhp,
    within=none,
]{panel}

\title{Pulsar timing array constraints on spin-2 ULDM}

\author[a]{Juan Manuel Armaleo,}
\author[a]{Diana L\'opez Nacir,}
\author[b]{and Federico R.~Urban}

\affiliation[a]{Departamento de Física  Juan Jos\'e Giambiagi, FCEyN UBA and IFIBA CONICET-UBA,
 Facultad de Ciencias Exactas y Naturales\\Ciudad Universitaria, Pabellon I, 1428 Buenos Aires, Argentina}
\affiliation[b]{CEICO, Institute of Physics of the Czech Academy of Sciences\\Na Slovance 2, 182 21 Praha 8, Czech Republic}

\abstract{Ultra-light Dark Matter (ULDM) models are suitable  candidates for  the cosmological Dark Matter that may leave characteristic imprints in many observables.  Among other probes, signatures of ULDM can be searched for in pulsar timing data.  In this work we describe the effects of spin-2 ULDM on pulsar timing arrays, extending previous results on lower spins.  Spin-2 ULDM is universally coupled to standard matter with dimensionless strength \(\al\).  We estimate that current data could constrain this coupling in the mass range \(m\lesssim4\times10^{-22}\)~eV at the \(10^{-5}\) to \(10^{-6}\) level, which is the most competitive constraint in this mass range.  A crucial feature of the spin-2 ULDM effect on pulsar timing is its anisotropic, quadrupolar shape.  This feature can be instrumental in differentiating the effects sourced by spin-2 ULDM from, for instance, scalar ULDM, and the systematics of a PTA experiment.}

\begin{document}
\maketitle
\flushbottom


\section{Introduction}
\label{sec:intro}

The existence of the cosmological Dark Matter (DM) is supported by a broad range of observations~\cite{Bertone:2016nfn}.  Even though its presence is (almost) undisputed, even the most fundamental properties of DM, such as its mass and interaction strength with Standard Model fields, are still mostly unknown (see for instance~\cite{Bertone:2018xtm} and references therein).  This ignorance derives from the fact that all experimental evidence for DM is purely gravitational, whereas attempts at using any other fundamental force to detect it have given null results (thus, they have only placed limits on interaction strengths).  In particular, the mass of the DM is largely  unconstrained, with viable candidates ranging from \(m\sim10^{-23}\)~eV to \(m\sim M_\odot\sim10^{66}\)~eV and beyond.

The lowest mass range, \(10^{-23}~\text{eV}\lesssim m \lesssim 10^{-17}\)~eV, usually dubbed Ultra-Light Dark Matter (ULDM), has been gaining a lot of traction in recent literature due to its peculiar properties at small scales, which markedly differentiate it from its heavier counterparts~\cite{Hui:2016ltb,Niemeyer:2019aqm,Hui:2020hbq}.  ULDM has been historically conceived in the form of scalar or pseudo-scalar fields such as axion-like particles or dilatons~\cite{Preskill:1982cy,Abbott:1982af,Dine:1982ah,Turner:1983he,10.1093/mnras/266.2.289,Hu:2000ke,Payez:2014xsa,Marsh:2015xka,Lee:2017qve,Ivanov:2018byi}.  However, nearly the same late Universe cosmological evolution and phenomenology can be obtained from a massive spin-1 or spin-2 field, as shown respectively in~\cite{Nelson:2011sf,Cembranos:2012kk,Cembranos:2012ng,Cembranos:2013cba,Arias:2012az,Graham:2015rva,Knapen:2017xzo} and~\cite{Marzola:2017,Aoki:2017cnz}.  The latter example is especially interesting because it arises directly as a \emph{modification of gravity} itself, even though it is in the guise of an additional particle, the DM. 

If DM is indeed ultra-light, at late times its behaviour can be approximated by a classical field which is rapidly oscillating with frequency \(m \gg H\), where \(H\) is the Hubble parameter.  Although the oscillations are negligible from a cosmological point of view, they can produce interesting effects in systems whose typical timescales are comparable to \(1/m\), see~\cite{Irsic:2017yje,Armengaud:2017nkf,Zhang:2017chj,Bullock:2017xww,Bar:2018acw,Robles:2018fur,Baryakhtar:2017ngi,Baumann:2018vus,Marsh:2018zyw,Nebrin:2018vqt,Safarzadeh:2019sre,Wasserman_2019,Khmelnitsky:2013lxt,Blas:2016ddr,LopezNacir:2018epg,Armaleo:2019gil,Blas:2019hxz}.

One such effect is the oscillation of the gravitational potentials along the line of sight of radio-loud pulsars, which leaves a mark in the times of arrival of the pulses.  Pulsar timing arrays (PTAs) are sensitive to this effect in the frequency range \(10^{-9}~\text{Hz}\lesssim\nu\lesssim10^{-6}\)~Hz, which corresponds to \(2\cdot10^{-24}~\text{eV}\lesssim m \lesssim2\cdot10^{-21}\)~eV.  PTAs have indeed been used to set competitive constraints on spin-0 and spin-1 ULDM models using this effect, see~\cite{Khmelnitsky:2013lxt,Nomura:2019cvc,Blas:2016ddr,LopezNacir:2018epg,Rozner:2019gba}.  In this work we focus instead on the spin-2 ULDM case~\cite{Armaleo:2019gil,Brito:2020lup}, which presents an interesting and unique phenomenology due to its tensorial structure as well as its action, which is non-negotiable at the peril of introducing ghosts.  This action unavoidably includes a direct coupling between the ULDM field and matter fields, as we will see below.  Our results are based on bimetric gravity but hold for any universally-coupled spin-2 ULDM.

The paper is structured as follows.  In section~\ref{sec:maths} we introduce the spin-2 field and work out the main effects of the oscillating field on the Earth-pulsar system. We focus there on the study of the ``Earth term'', which is the dominant part of the signal. In section~\ref{sec:res}, we present and discuss our results, and the constraints that we can obtain from existing data.  In section~\ref{sec:out} we conclude with a summary of our results and an outlook for future studies.  Appendix~\ref{appendix} contains the generalisation of the calculation of section~\ref{sec:maths} to include additional contributions to the signal.

\section{PTA frequency shift for spin-2 ULDM}
\label{sec:maths}

\subsection{The spin-2 ULDM field}\label{ssec:uldm}

We consider a massive spin-2 field \( \M_{\mu\nu} \) described by the Fierz-Pauli lagrangian density
\begin{align}\label{PF}
  {\cal L} &\, \deq \frac12 \M_{\mu\nu}\mathcal{E}^{\mu\nu\rh\si}\M_{\rh\si} - \frac14 m^2 \left( \M_{\mu\nu}\M^{\mu\nu} - \M^2 \right) \,, 
\end{align}
where \( \M \deq g^{\mu\nu} \M_{\mu\nu} \), and the Lichnerowicz operator \( \mathcal{E}^{\mu\nu\rh\si} \) is defined by
\begin{align}
  \mathcal{E}^{\mu\nu}_{~~\rh\si} \deq &\, \de^\mu_\rh \de^\nu_\si \Box - g^{\mu\nu} g_{\rh\si} \Box + g^{\mu\nu} \D_\rh \D_\si + \nn \\
  &\, + g_{\rh\si} \D^\mu \D^\nu - \de^\mu_\si \D^\nu \D_\rh - \de^\mu_\rh \D^\nu \D_\si \,. \label{eq:lich}
\end{align} 
For a  Friedman-Lema\^itre-Robertson-Walker (FLRW) background metric,  
the equations of motion for the ULDM field  in the late-time universe,  when \( m\gg  H \)  (with  \(H\)  the Hubble rate), can be derived following ~\cite{Marzola:2017,Aoki:2017cnz}. From the
 Bianchi identities one can identify the five propagating  degrees of freedom of  \( \M_{\mu\nu} \), which can  be described by the six \( \Mij \) components, subject to the additional tracelessness constraint \( \M^i_{~i}=0 \). These  components satisfy 
\begin{align}
  \ddot\M_{ij} + 3H\dot\M_{ij} -\bigtriangleup \Mij+ m^2\Mij &\, = 0 \,, \label{eq:eom}
\end{align}
where \(\bigtriangleup\) is the spatial Laplace operator.  The homogeneous background solution is given by
\begin{align} 
  \Mij &\, = \frac{\hat{\M}_{ij}}{ R^{3/2}} \cos{(mt+\Upsilon)}\vep_{ij}  \,, \label{eq:hij}
\end{align}
where \(R\) is the FLRW scale factor.  The overall amplitude $\hat{\M}_{ij}$ is fixed so that the ULDM energy density matches the observed  background DM density,  and \(\vep_{ij} (\vx)\) is an angular  matrix with unit norm, zero trace and is symmetric (see~\cite{Maggiore:1900zz} and Appendix A of~\cite{Armaleo:2019gil}).   With this solution, it can be  shown that   the background ULDM field behaves as a  suitable  DM candidate, with an   energy density scaling as \(R^{-3}\), and a pressure that averages to zero on the large time-scales relevant for the cosmological background evolution.  

On the astrophysical scales we are interested in we can set \(R=1\).  The local value of the ULDM field at the position \(\vx\) can be described by the oscillating function~\cite{Marzola:2017,Armaleo:2019gil}: 
\begin{align}\label{eq:mij}
	\Mij & = \frac{\sqrt{2\rhoDM(\vx)}}{m}\cos{(mt+\Upsilon(\vx))}\vep_{ij}(\vx) \,, 
\end{align}
where \(\rhoDM (\vx)\) is the observed local dark matter energy density, for which we assume the conservative value of \(\rhoDM=0.3\)~GeV/cm\(^3\) \cite{Piffl:2014mfa,Evans:2018bqy,2015ApJ...814...13M,pdg2020}, \( \Upsilon(\vx) \) is a random phase. The spatial gradients of the  field are expected  to be relevant at scales of order of the de~Broglie wavelength \(\ldB \deq 2\pi(m V)^{-1}\), where \(V\) is the effective velocity of the ULDM, which in the galactic halo can be estimated from the virial velocity. For the Milky Way halo we assume \(V_0\sim 10^{-3}\). Therefore, we can estimate
\begin{align}
	\ldB & \sim 4\, {\rm  kpc}\left(\frac{10^{-3}}{V}\right)\left(\frac{10^{-23}{\rm eV}}{m}\right).
\end{align}
If the characteristic scale of the inhomogeneities of the DM field is given by \(\ldB^{-1}\), the gradients of the field can be important for PTAs, because the distances between the Earth and the pulsars can be of the same order or even larger.  However, in the situation we consider below, the main contribution to the signal depends only on the field configuration near the Earth, where it is a reasonable assumption that gradient effects can be treated perturbatively.  Notice that, because \(\partial^{\mu}{M}_{\mu\nu} = 0\), the components \(M_{0i}\) (\(M_{00}\)) are of first (second) order in gradients of \(\Mij\).

In what follows we neglect second order gradient effects.  Nevertheless, in Appendix~\ref{appendix} we discuss on the possibility of performing a more dedicated study, including the contribution of all \({M}_{\mu\nu}\) components to the signal.  Such a study necessarily requires further assumptions and modelling of the field inhomogeneities.  Lastly, notice that \Eq{eq:mij} assumes a constant common frequency.  This applies as long as the field remains coherent in time. The coherence time is expected to be given by
\begin{align}
	t_{\rm coh} &\deq \frac{\pi}{m V^2}\sim 6.5 \times 10^6\, {\rm yr}\left(\frac{10^{-3}}{V}\right)^2\left(\frac{10^{-23}{\rm eV}}{m}\right).
\end{align}
In this work we consider sufficiently light fields for which \(t_{\rm coh}\) is much longer than the observation time-scale, and therefore the field remains coherent with a time dependence given by \Eq{eq:mij}.

\subsection{The Earth-pulsar system}\label{ssec:sys}

The Earth-pulsar system in presence of the spin-2 ULDM is described by the action
\begin{align}\label{eq:action}
	S &\deq S_\text{free}[g,\Mmn,\Psi] + S_\text{int}[g,\Mmn,\Psi] \,,
\end{align}
where the first piece \(S_\text{free}[g,\Mmn,\Psi]\) represents the free action of matter, denoted by \(\Psi\), which in our case are photons propagating from pulsars to Earth along the null geodesics of the space-time metric \(g_{\mu\nu}\).  This action includes the quadratic Fierz-Pauli lagrangian for the ULDM field given in ~\Eq{PF}.  The second piece describes the interaction between the ULDM field \(\Mmn\) and the system (including photons), and is given by
\begin{align}\label{eq:int}
  S_\text{int}[g,\Mmn,\Psi] & \deq -\frac{\al}{2\mpl} \int\!\dd^4x\,\sqrt{-g} \Mmn T_\Psi^{\mu\nu} \,,
\end{align}
where \(\al\) is the strength of the interaction, \(\mpl\) is the reduced Planck mass, and \(T_\Psi^{\mu\nu}\) is the energy-momentum tensor of the free system.  This interaction term is idiosyncratic for spin-2 ULDM because it is required by the self-consistency of the model; in particular, there is no ULDM at all with \(\al\rar0\) because the ULDM field becomes infinitely strongly coupled in this limit, and decouples completely from all other fields, even gravitationally.  This is markedly different in spin-0 and spin-1 ULDM models, where a direct interaction term (a fifth force) has to be included by hand.


In order to obtain the effect of the spin-2 ULDM on the Earth-pulsar system we begin by noticing that we can do away with the interaction term altogether by changing the frame according to
\begin{align}\label{eq:coord}
    \tilde{g}_{\mu\nu} & \deq g_{\mu\nu} + \frac{\al}{\mpl}\Mmn \,.
\end{align}
In this frame the Earth, the pulsars, and photons no longer interact directly with the ULDM.  However, photons travelling from the pulsar to the Earth will follow the geodesics of the new metric \(\tilde{g}\), which explicitly depends on the ULDM field \(\Mmn\).

To see this, we observe that  the interaction term remains the same because it is already first order in \(\alpha\).  The free action instead gives
\begin{align}\label{eq:trick}
  S_\text{free}[g,\Mmn,\Psi] & \simeq S_\text{free}[\tilde{g},\Mmn,\Psi] + \frac{\al}{2\mpl} \int\!\dd^4x\,\sqrt{-\tilde{g}} \Mmn T_\Psi^{\mu\nu} \,.
\end{align}
In other words, it is possible to ``reabsorb'' the massive spin-2 field into the metric, as is expected from the original formulation of bimetric gravity~\cite{Hassan:2011zd}\footnote{Here we are keeping only the terms that produce an effect on the pulsar systems (including the photons) that are linear in \(\alpha\).  Notice in particular that the cubic self-interaction of the ULDM field arising when the transformation in \Eq{eq:coord} is applied to the Fierz-Pauli action does not change the result.}, owing to the fact that the coupling between the ULDM and matter fields is \emph{universal}; in the case of bimetric gravity this is dictated by the spin-2 nature of the ULDM field.  Therefore, the final action of the system in the \(\tilde{g}\) frame is simply
\begin{align}\label{eq:newframe}
	S &= S_\text{free}[\tilde{g},\Mmn,\Psi] \,.
\end{align}

\subsection{Time residuals: the Earth term}\label{ssec:freq}

In this section we compute the main ULDM effect on the time residual of radio pulses (see Appendix~\ref{appendix} for a generalisation).  We work in the frame in which photons travel along geodesics of the metric \(\tilde{g}_{\mu\nu}\) defined in \Eq{eq:coord}.  The contribution of the \(M_{0\mu}\) components to the signal we consider below are of second order and higher in a derivative expansion of the field \(M_{ij}\) near the Earth, so they can be neglected. 
Therefore, we set \(M_{0\mu}=0\) and work with
\begin{align}\label{eq:metric}
	\tilde{g}_{ij} &= -\delta_{ij} + \frac{\al}{\mpl}\Mij \,.
\end{align}

Let us consider a photon with unperturbed four-momentum \(p^\mu\deq(\nu,\nu n^i)\), frequency \(\nu\) and momentum along the unit three-vector \(\vn\deq n^i\).  The change in frequency along the geodesics  
is given by
\begin{align}\label{eq:geo}
	\frac{\dd p^0}{\dd s} & = -\Gamma^0_{ij} \, p^i p^j = \frac{\al\nu^2}{2\mpl} \, \partial_0 \Mij \, n^i n^j \,,
\end{align}
where \(s\) is the affine parameter and \(p^0 = \dd x^0/\dd s\).  Then, keeping only the linear terms in \(\al\),
\begin{align}\label{eq:freq0}
	\nu & = \nu_0  \left\{1+ \frac{\al}{2\mpl} \int\limits_\ostar^\oplus\!\dd s\,\nu_0 \pd_0\Mij \, n^i n^j  \right\} \,,
\end{align} 
where \(\nu_0\) is the unperturbed frequency at the pulsar.  Using that \(\nu_0 \partial_0=\frac{d}{d s}-n^i \nu_0\partial_i\), we obtain
\begin{align}\label{eq:freq02}
	\nu & = \nu_0  \left\{1+ \frac{\al}{2\mpl} \left(\Mij^\oplus-\Mij^\ostar\right)\, n^i n^j  -\frac{\al}{2\mpl} \int\limits_\ostar^\oplus\!\dd s\,\nu_0 n^l\partial_l \Mij \, n^i n^j  \right\} \,,
\end{align} 
where the index \(\oplus\) (\(\ostar\)) on a quantity means it is to be evaluated at the Earth (pulsar).

The distance to a pulsar is typically much larger than the Compton wavelength of the ULDM (for example, \(m\sim10^{-23}~\text{eV}\sim10^{-8}~\text{Hz}\sim1/\text{pc}\)).  However, as we mentioned above, the distance between the Earth and a pulsar can be comparable to the de~Broglie wavelength, so the  ULDM field configuration is in principle different on Earth and at the location of the pulsar.  Hence, the highly oscillating integral in \Eq{eq:freq02} is subdominant with respect to the first two terms.  Roughly, this term is suppressed by a factor \(v/c\sim\mathcal{O}(10^{-3})\) coming from the gradient \(\pd_l \Mij\).  The terms evaluated at the Earth (\(\oplus\)) and at the pulsar (\(\ostar\)) are, respectively, the so-called Earth and pulsar terms~\cite{Porayko:2014rfa,Porayko:2018sfa}.   When studying the cross-correlations between the different signals coming from many pulsars, for a  frequency bin given by the ULDM mass, the result can be split into the contribution of the pulsar terms, the Earth terms and the combined pulsar-Earth terms.  Notice that in our case, not only the phases \(\Upsilon\) as in the scalar case~\cite{Porayko:2018sfa,DeMartino:2017qsa,Blas:2016ddr,Khmelnitsky:2013lxt} but also the geometry of a given realisation of the ULDM quadrupole is generically not the same on Earth and at the location of each pulsar.  In other words, the orientation of the two multipole vectors\footnote{Multipole vectors were introduced in~\cite{Copi:2003kt} as an alternative way to parametrise and visualise spherical harmonic coefficients on a sphere.} that define the ULDM quadrupole is in general different on Earth than at the pulsars.  For this reason, the main contribution to the cross-correlations will be given by the Earth term, because the effect of the ULDM field on the photon path is expected to sum up coherently for the end of the path, i.e., near the Earth, whereas the other terms are expected to wash out on average.  Therefore, the Earth term dominates the residuals  and we keep only this term in what follows---see Appendix~\ref{appendix} for further discussion.

Plugging the expression for the DM field \Eq{eq:mij} into \Eq{eq:freq02}, the Earth term becomes
\begin{align}\label{eq:freq}
	\nu
	& = \nu_0 \left\{ 1+ \frac{\al}{\sqrt2 m\mpl}  \sqrt{\rhoDM}_{\oplus}\vep_{ij,\oplus}\cos\left(mt+\Upsilon_\oplus\right) n^in^j  \right\} \,.
\end{align} 
This frequency shift \Eq{eq:freq} induces a time residual in the radio pulses given by
\begin{align}\label{eq:det}
	\trt & \deq -\int\limits_0^t\!\dd t' \, \frac{\nu-\nu_0}{\nu_0} \,.
\end{align}
Subtracting the average time residual over the observation time---PTAs are sensitive only to the time variation of the residual---we obtain
\begin{align}\label{eq:shift_earth}
	\trt & = -\frac{\al\sqrt{\rhoDM}_{\oplus}}{\sqrt2m^2\mpl} \vep_{ij,\oplus}n^in^j \sin\left(mt+\Upsilon_\oplus\right) \,.
\end{align}
This is the main analytical result of the paper; we will proceed to use this expression to constrain spin-2 ULDM in the following section.

\section{Results and discussion}\label{sec:res}

\subsection{The geometry of the time residuals}\label{ssec:geo}

The ULDM spin-2 field is described by five different polarisation states (helicities).  Because we have already normalised the overall amplitude to match the observed dark matter energy density, we can parametrise the spin-2 field with three amplitudes \(\vepS\), \(\vepV\), and \(\vepT\) that obey \(\vepS^2+\vepV^2+\vepT^2=1\), and two angles \(\eta\) and \(\chi\), see   Appendix A of~\cite{Armaleo:2019gil},
\begin{align}\label{eq:quad}
  \vep_{ij} & = \frac{1}{\sqrt2}\left(
  \begin{array}{ccc}
    \vepT \cos\chi - \vepS/\sqrt3 & \vepT \sin\chi & \vepV \cos\eta \\
    \vepT \sin\chi & - \vepT \cos\chi - \vepS/\sqrt3 & \vepV \sin\eta \\
    \vepV \cos\eta & \vepV \sin\eta & 2\vepS/\sqrt3 \\
  \end{array} \right) \,.
\end{align}
The angle \(\eta\) determines the azimuthal direction of the ``vector'' helicities of the spin-2 field; the angle \(\chi\) instead determines the azimuthal orientation of the ``tensor'' helicities of the spin-2.

The global shape of the quadrupolar time residual can be visualised in a spherical harmonic representation \(\vep_{ij}n^in^j \deq \sum_m a_m Y^{2m} \) where \( Y^{2m}(\vn) \) are the real spherical harmonics for the \( \vn \deq (x,y,z) \deq (\sin\vth\cos\vph,\sin\vth\sin\vph,\cos\vth)\) unit coordinate vector.  We normalise the spherical harmonics as
\begin{align}
  & Y^{2,-2} = \sqrt2 xy \,, & Y^{2,2} &= \left(x^2-y^2\right) / \sqrt2 \,, & \nn \\
  & Y^{2,-1} = \sqrt2 yz \,, & Y^{2,1} &= \sqrt2 xz \,, & \nn \\
  & Y^{2,0} = \left(3z^2-1\right) / \sqrt6 \,. &&&
\end{align}
The spherical harmonic coefficients are then
\begin{align}
  & a_{-2} = \vepT \sin\chi \,,\quad a_{2} = \vepT \cos\chi \,, \nn \\
  & a_{-1} = \vepV \sin\eta \,,\quad a_{1} = \vepV \cos\eta \,, \nn \\
  & a_{0} = \vepS \,.
\end{align}

\subsection{Estimating the limits}\label{ssec:plots}
To make contact and exemplify the differences with the known results for spin-0 ULDM and a stochastic gravitational wave background, we average \Eq{eq:shift_earth} over the celestial sphere \(\vn\) to obtain
\begin{align}\label{eq:shift_earth_avg}
	\trSq &= \frac{\al\sqrt{\rhoDM}_{\oplus}}{\sqrt{15}m^2\mpl} \sin\left(mt+\Upsilon_\oplus\right) \,.
\end{align}
Note that, whereas the frequency shift for any given pulsar, \Eq{eq:shift_earth}, depends on all five parameters describing the quadrupole, in general the average over the sphere can only depend on up to three of them, because we can choose the coordinate system as we wish (for example, \(\chi=0=\eta\))\footnote{Other choices are possible, but not all of them are general: for example the axisymmetric configuration with \(\vepV=0=\vepT\) does not represent a generic quadrupole, see for instance~\cite{Thorsrud:2013kya,Ramazanov:2016gjl}.}.  However, owing to the symmetry of the system, the result \Eq{eq:shift_earth_avg} depends only on the overall amplitude of the quadrupole, as set by \(\al\), whereas the other parameters describing the quadrupole have dropped out.

We can now compare \Eq{eq:shift_earth_avg} with the time residual caused by an incoming train of background stochastic gravitational waves of frequency \(\omega\) and strain \(h_c\)~\cite{Wen:2011xc}:
\begin{align}\label{eq:shift_gw}
	\trSq &= \frac{h_c}{\sqrt6\omega} \sin\left(\omega t+\Upsilon_\oplus\right) \,,
\end{align}
we see how the spin-2 ULDM corresponds to a strain
\begin{align}\label{eq:strain_avg}
	h_c &= \frac{\al\sqrt{2\rhoDM}}{\sqrt{5}m\mpl} \,,
\end{align}
where we have taken \(m=\omega\).

Notice that the frequency of the oscillations of the time residuals is given by \(m\), not by \(2m\) as in the case of the indirect effect on the gravitational potentials of the spin-0 and spin-1 ULDM models, see~\cite{Khmelnitsky:2013lxt,Nomura:2019cvc,Blas:2016ddr}.  This is because the ULDM field in that case enters through its energy-momentum tensor, which is quadratic in the field.  In the direct coupling case instead the contribution is already at the linear level, see \Eq{eq:metric}.  This is also the reason why a spin-1 ULDM would produce an effect on PTA whose angular dependence corresponds to \(Y^{2,0}(\vn)\) instead of \(Y^{1,0}(\vn)\), which is expected if the spin-1 field is active at the linear level already.

\begin{figure}
\centering
	\includegraphics[width=0.8\textwidth]{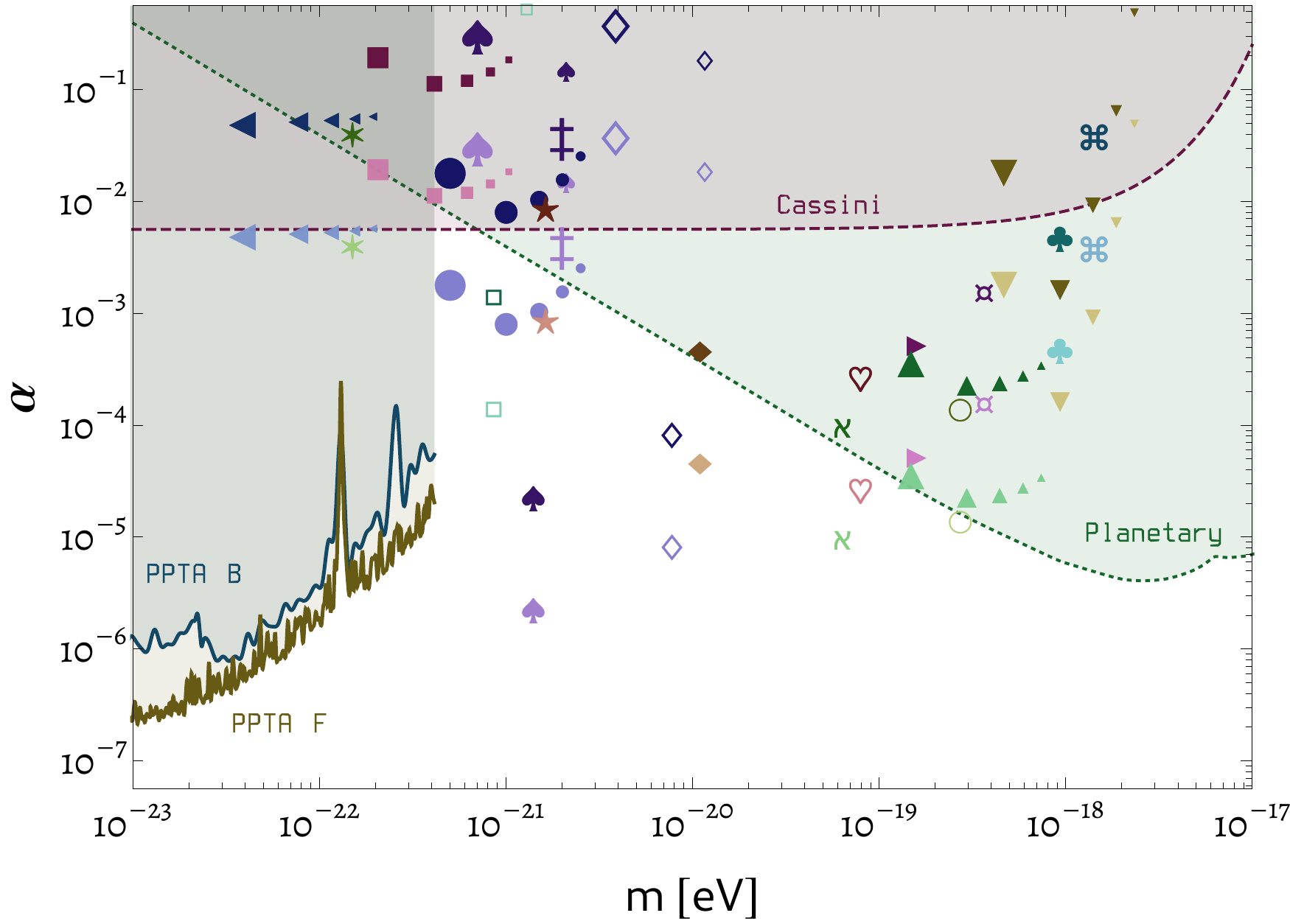}
	\caption{PPTA bounds on the strength of the spin-2 ULDM coupling \(\al\) versus the ULDM mass \( m \); we show here the results of the bayesian (PPTA B) and frequentist (PPTA F) analyses, reproduced with permission from~\cite{Porayko:2018sfa}.  The shaded region above the dashed purple line is excluded by solar system tests~\cite{Hohmann:2017uxe}.  The shaded region above the dotted green line is excluded by planetary constraints~\cite{Sereno:2006mw}.  Each symbol represents the constraint derived from the non-observation of secular variations in the orbital parameters of a binary pulsar system, see~\cite{Armaleo:2019gil}.}\label{fig:alpha}
\end{figure}

In figure~\ref{fig:alpha} we collect the bounds on the strength of the spin-2 ULDM coupling \(\al\) versus the mass \(m\).  The level of the bounds that can be obtained with PPTA are taken form the bayesian (PPTA B) and frequentist (PPTA F) analyses of~\cite{Porayko:2018sfa}.  The limits we show are indicative of what can be done with PPTA, but are not precise bounds since we are comparing the all-sky average \Eq{eq:shift_earth_avg}, which does not account for the specificities of each pulsar, to the PPTA results.  Other limits included in this figure are: the dashed purple line is the limit from solar system tests~\cite{Hohmann:2017uxe}; the dotted green line is the limit obtained from planetary constraints~\cite{Sereno:2006mw}; the symbols represent the constraints derived from the non-observation of secular variations in the orbital parameters of a binary pulsar system, see~\cite{Armaleo:2019gil} for a detailed explanation---the limits obtained in this case are only comparable at the order of magnitude level because we do not know the ULDM configuration for each binary pulsar system.

\begin{figure}
\centering
	\includegraphics[width=0.8\textwidth]{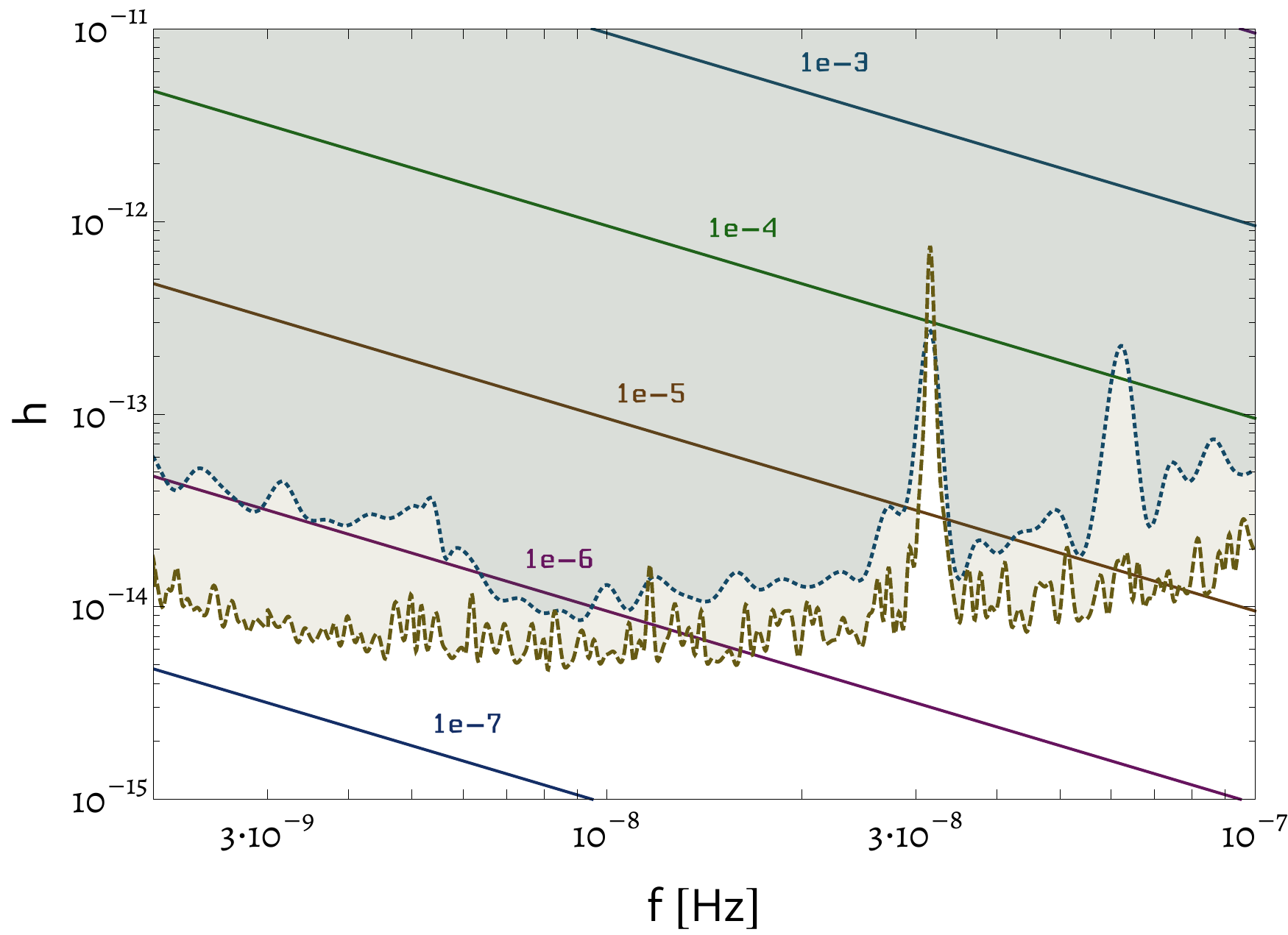}
	\caption{PPTA bayesian (blue, dotted) and frequentist (yellow, dashed) bounds on the equivalent gravitational wave strain as a function of frequency, reproduced with permission from~\cite{Porayko:2018sfa}.  Also shown are the equivalent strain produced by spin-2 ULDM as calculated in eq.~(\ref{eq:strain_avg}), for several values of the ULDM coupling \(\al\).}\label{fig:strain}
\end{figure}  

In figure~\ref{fig:strain} we present once again the PPTA bayesian (blue, dotted) and frequentist (yellow, dashed) bounds on the equivalent gravitational wave strain as a function of frequency, from~\cite{Porayko:2018sfa}.  The average equivalent strain produced by spin-2 ULDM, as calculated in eq.~(\ref{eq:strain_avg}), is shown for several values of the ULDM coupling from \(\al=10^{-7}\) to \(\al=10^{-3}\).  Only the lowest values, depending on the frequency, are allowed, as shown in figure~\ref{fig:alpha}.

The ULDM field is also going to contribute with an oscillatory term to the gravitational potentials (and therefore the equivalent strain) as in the spin-0 case~\cite{Khmelnitsky:2013lxt}.  This countribution however is going to be subdominant and, in terms of the effective coupling \(\al_\text{eff}\sim\sqrt{\rhoDM}/m\mpl\) is, for the smallest \(m\sim10^{-23}\)~eV, of order \(\al_\text{eff}\sim10^{-7}\).

\subsection{Correlation}\label{ssec:HD}

By studying the correlation between different signals, one may clarify some aspects of the nature of DM (see~\cite{Ramani:2020hdo} for an application to small-scale ULDM structures).  Here we study the anisotropies of the ULDM signal in the spin-2 case through the correlation between the response functions of pairs of pulsars, analogously to the Hellings and Downs curve for searches for gravitational wave backgrounds~\cite{HDoriginal} (for some pedagogical references see~\cite{Mingarelli:2016zug,Jenet:2014bea}).  The response function for each pulsar is given by \Eq {eq:shift_earth}.  Therefore, the correlation function \(C(\vth,\vph)\deq\tr^a(t_a)\,\tr^b(t_b) \) for two pulsars \(a\) and \(b\) located at \(\vn_a\) and \(\vn_b\) and observed at time \(t_a\) and \(t_b\), respectively, is simply:
\begin{align}\label{eq:corr_gen}
    C(\vth,\vph) = \, & \frac{\al^2\rhoDM}{2m^4\mpl^2} \sin\left(mt_a+\Upsilon_\oplus\right) \sin\left(mt_b+\Upsilon_\oplus\right) \vep_{ij,\oplus} \, \vep_{kl,\oplus} \, n^i_a n^j_a n^k_b n^l_b \,.
\end{align}

We can choose the coordinate system by aligning pulsar \(a\) with the \(z\)-axis:
\begin{align}\label{eq:frame}
    & \vn_a=(0,0,1) \,, \quad \vn_b=(\sin\vth\cos\vph,\sin\vth\sin\vph,\cos\vth) \,.
\end{align}

Note, however, that with our parametrisation in \Eq{eq:quad} we are separating the quadrupole components according to their properties under rotations around the \(z\)-axis; therefore, with the choice of coordinates \Eq{eq:frame} only the scalar helicity will contribute to the final result from the pulsar \(a\).  This choice is nevertheless useful to make contact with the literature discussing the signal expected from other effects, such as a gravitational wave background in General Relativity and its extensions~\cite{Jenet:2005pv,Lee:2008,Lee:2010cg,Chamberlin_2012,Gair:2015hra}.  The correlation function reads
\begin{align}\label{eq:corr}
	C(\vth,\vph) = \,	& \frac{\al^2\rhoDM}{6m^4\mpl^2} \sin\left(mt_a+\Upsilon_\oplus\right) \sin\left(mt_b+\Upsilon_\oplus\right) \times \nn\\
				\times	& ~\vepS \left\{ \vepS \left[3\cos^2\vth-1\right] + \sqrt3\left[ \vepV \sin2\vth \cos(\vph-\eta) + \vepT \sin^2\vth \cos(2\vph-\chi) \right] \right\} \,.
\end{align}

\begin{figure}
\centering
	\includegraphics[width=0.9\textwidth]{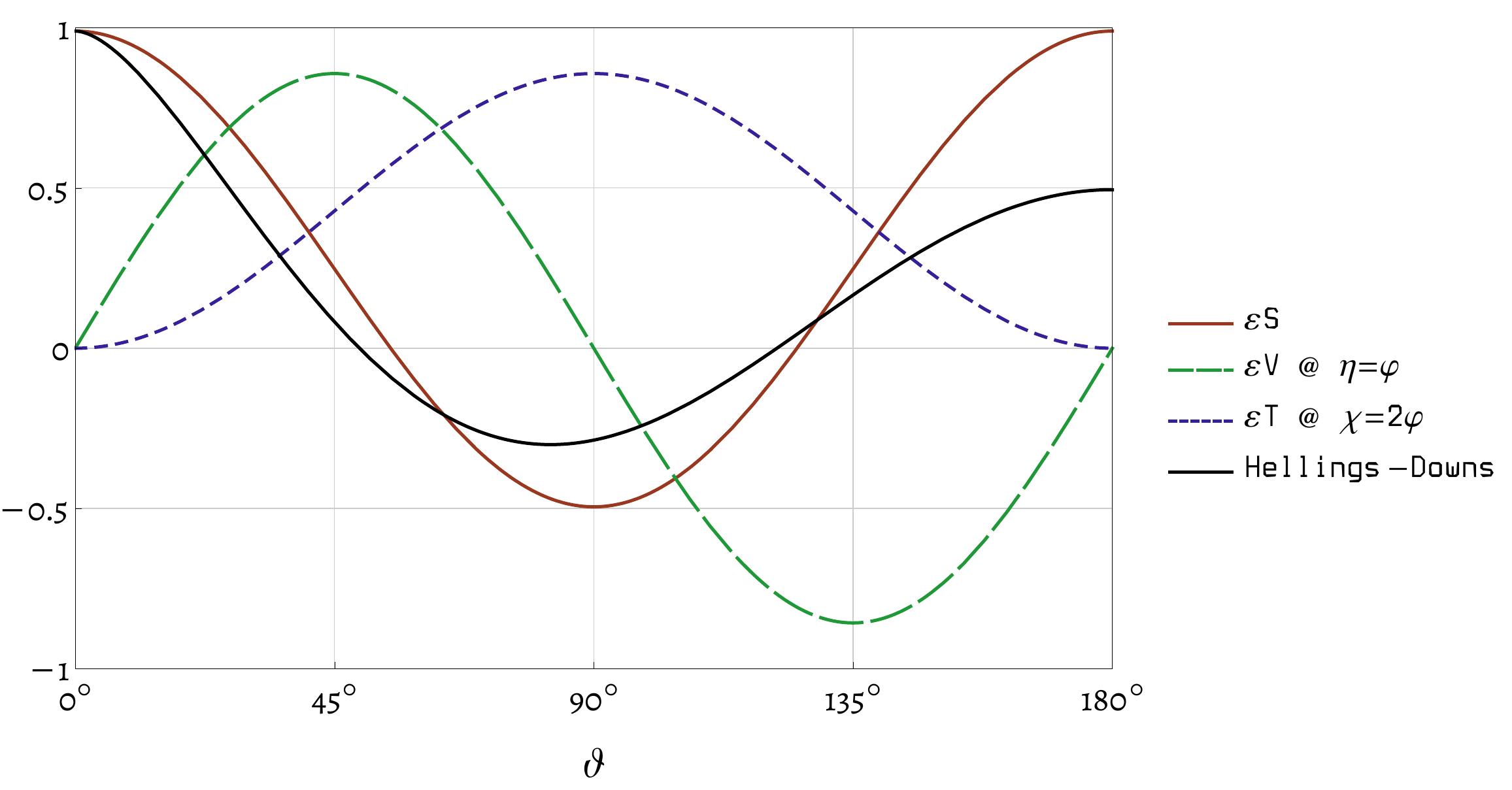}
	\caption{The polar angle dependence of the correlation function \(C(\vth,0)\) of \Eq{eq:corr} for the three helicities \(\vepS\) (red, solid), \(\vepV\) (green, long dashes), and \(\vepT\) (purple, short dashes), normalised such that \(\vepS\) contributes 1 when \(\vth=0\).  The correlation contribution of the scalar helicity \(\vepS\) clearly does not depend on the azimuthal angle \(\vph\).  We have plotted the vector helicity contribution aligned with \(\vph=\eta\), and the tensor one, \(\vepT\), aligned with \(2\vph=\chi\). Also, for reference, we show the Hellings and Downs curve; the relative normalisation between the Hellings and Downs curve and the three contributions from the three spin-2 helicities is arbitrary.}\label{fig:corr}
\end{figure}

We show the polar behaviour of the correlation function for the three helicities in figure~\ref{fig:corr}, keeping the caveat of the choice of coordinate system in mind.  Of course in reality we will see the sum of all three, and in order to disentangle them one needs to further look at the azimuthal dependence of the signal: the scalar term is independent of \(\vph\), the vector helicity behaves as \(\cos(\vph-\eta)\), whereas the tensor one is proportional to \(\cos(2\vph-\chi)\).

\section{Conclusion and outlook}\label{sec:out}

In this work we have studied the effects of spin-2 ULDM on pulsar timing for a PTA setup.  The cosmological late-time ULDM oscillatory behaviour directly induces a characteristic time-dependent and direction-dependent shift in the frequency of the radio waves emitted by pulsars, \Eq{eq:shift_earth}.  This is so because spin-2 ULDM is universally coupled to the energy-momentum tensor of standard matter, a direct coupling that is parametrised by the constant \(\al\), see~\Eq{eq:int}.

An important feature that distinguishes the spin-2 ULDM effect on pulsar timing from other sources of time residuals is its anisotropy: the magnitude of the time residuals depends on the pulsars' positions in the sky.  This dependence is quadrupolar, that is, for each of the five degrees of freedom of the spin-2 ULDM field we can associate a spherical harmonic of degree 2, which describes the strength of the effect as we move around on the celestial sphere, see section~\ref{ssec:geo}.

We have shown how, for the lowest mass range of interest for ULDM models, namely \(m\lesssim4\times10^{-22}\)~eV, existing data from, e.g., PPTA, can lead to competitive constraints on the spin-2 ULDM coupling strength, figures~\ref{fig:alpha} and~\ref{fig:strain}.  Our results are an estimation of the level of the constraints that can be obtained with current PTA data, for which we have averaged the ULDM effects over the sphere in~\Eq{eq:shift_earth_avg}; the actual analysis along the lines of~\cite{Porayko:2014rfa} would be able to take advantage of the anisotropy in the signal to optimise the constraints and to differentiate this effect from, e.g., a spin-0 ULDM model.

In section~\ref{ssec:HD} we have obtained the correlation between time residuals from any two pulsars in the array, \Eq{eq:corr}.  On account of the quadrupolar nature of the spin-2 ULDM field, this correlation explicitly depends on the angular separation between the pair of pulsars, as well as their relative azimuthal position.  This is a peculiar feature that not only sets this model apart from the signatures of ULDM models, but also has important implications for the search strategies for this signal.  Indeed, different systematic effects in PTA studies for correlations among pulsar pairs can be separated, and therefore dealt with in the analysis, thanks to, among other features, their different anisotropic behaviours~\cite{Verbiest:2018ysd,Taylor:2016gpq,Tiburzi:2015kqa}.  For example, a systematic error in the clock time standard would be monopolar (i.e., isotropic), and a systematic error in the planetary ephemeris would be dipolar.

We conclude with an outlook for future analyses.  The spin-2 ULDM effects on pulsar timing are similar to those generated by a steady and distant source of monochromatic gravitational waves, for example from a super-massive black hole binary system in the early stage of coalescence.  Several methods and actual searches for this signal exist both in the time domain~\cite{Lee:2011et,Zhu:2014rta,Madison:2015txa}, and frequency domain~\cite{Zhu:2015tua}.  It would be interesting to see to which extent these methods can be applied to the spin-2 ULDM case we have discussed here, and how well the two types of signals can be separately identified or could be mistaken one for the other.

\acknowledgments

The Authors would like to thank N.~K.~Porayko and Z.~Zhu for providing the PPTA constraints of~\cite{Porayko:2018sfa}.  FU is supported by the European Regional Development Fund (ESIF/ERDF) and the Czech Ministry of Education, Youth and Sports (MEYS) through Project CoGraDS - \verb|CZ.02.1.01/0.0/0.0/15_003/0000437|.  The work of DNL and JMA has been supported by CONICET, ANPCyT and UBA.

\appendix
\section{Beyond the Earth term}\label{appendix}
In this Appendix we generalise the calculation of the ULDM effect on the time residual presented in section~\ref{ssec:freq}.  As assumed in~\cite{Armaleo:2019gil} for the spin-2 case and as is normally done for scalar ULDM  models (see, for instance,~\cite{Khmelnitsky:2013lxt,Hui:2016ltb,Blas:2016ddr,Blas:2019hxz,Hui:2020hbq}), inside the Milky Way halo, the DM field can be taken to be homogeneous over spatial regions much smaller than a patch of characteristic size of order of the de~Broglie scale (a coherent dB patch), with inhomogeneties appearing on larger scales.  Therefore, studying the correlation between the different signals coming from pulsars that are not in the same coherent dB patch, beyond the Earth term, requires a model for the ULDM field correlation across different patches.  For instance, in the scalar case, one could consider a random phase model (see for example~\cite{Derevianko:2016vpm,Foster:2017hbq,Centers:2019dyn,Hui:2020hbq}).  Moreover, for the spin-2 case, a model for the behaviour of the polarisation tensor is required.  We discuss some of these aspects below.

The local spin-2 ULDM field given in \Eq{eq:mij} is a particular solution to the wave-equation 
\begin{align}\label{Fieldeq}
	& \Box \Mmn + m^2\Mmn=0 \,,
\end{align}
subject to the following  constrains: \(M=M^{\mu}_{~\mu}=0\) and \(\partial^{\mu}\Mmn=0\).  As mentioned above, for a FLRW  background,  this equation can be derived from the bimetric gravity action if one assumes \(m\gg H\), where \(H\) is the Hubble rate. As  shown in~\cite{Marzola:2017},  the derivation does not require  any further assumptions about the magnitude of spatial derivatives of \(\Mmn\).  Moreover, in the regime \(m\gg H\), this is the generic equation of a spin-2 field.

Given the metric \(\tilde{g}_{\mu\nu}\) defined in \Eq{eq:coord}, we can write a general formula for the frequency shift \Eq{eq:freq0} where all components of \(\Mmn\) are taken into account.  The geodesic equation is then
\begin{align}\label{eq:geo_gral}
	\frac{\dd p^0}{\dd s} & =-\Gamma^0_{\mu\nu}p^\mu p^{\nu} \nn\\
	& =-\frac{\nu^2\al}{2\mpl} \left\{ \pd_0M_{00}+2\pd_i M_{00} n^i + \left(\pd_jM_{0i}+\pd_iM_{0j}-\pd_0M_{ij}\right) n^in^j\right\} \,,
\end{align}
where we have kept only first-order terms in the (small) coupling constant \(\al\). Taking into account the local redshift factor \( 1/\sqrt{\tilde{g}_{00}}\) associated to the perturbation of clocks on Earth and the pulsar period, the generalised frequency shift becomes
\begin{align}\label{eq:freq0_gral}
	\nu = \nu_0 \Bigg\{1 & +\frac{\al}{2\mpl} \left(M_{00}^\oplus-M_{00}^\ostar\right) \nn\\
	& \left. -\frac{\al }{2\mpl} \int\limits_\ostar^\oplus\! \dd {s}\, \nu_0\left[\pd_0M_{00} + 2\pd_iM_{00}n^i + \left(\pd_jM_{0i}+\pd_iM_{0j}-\pd_0M_{ij}\right) n^in^j\right] \right\} \,.
\end{align}
Now using \(\nu_0 \pd_0=\frac{d}{d s}-\nu_0n^{i}\pd_i\) to rewrite the first and the last term inside the integral, we find
\begin{align}\label{eq:freq0_gral2}
	\nu = \nu_0 \Bigg\{1 & +\frac{\al}{2\mpl}\left(M_{ij}^\oplus-M_{ij}^\ostar\right)n^in^j \nn\\
	& \left. -\frac{\al }{2\mpl} \int\limits_\ostar^\oplus \! \dd {s} \,\nu_0\left[n^l\pd_l(M_{00}+n^i n^jM_{ij}) + \left(\pd_jM_{0i}+\pd_iM_{0j}\right) n^in^j\right] \right\} \,.
\end{align}
Taking into account the constraint \(\pd^\mu \Mmn=0\) we obtain
\begin{align}
	M_{0i} & = \pd_j\int \! \dd t' M_{ij}(t',\vx) + K_{0i}(\vx)\,, \\
	M_{00} & = \pd_i\int \! \dd t' M_{0i}(t',\vx) = \pd_i\pd_j \int \! \dd t' \int^{t'} \! \dd t'' M_{ij}(t'',\vx) + \pd_i K_{0i}(\vx) t + K_{00}(\vx) \,.
\end{align}
Then it is clear that the contribution of \(M_{00}\) and \(M_{0i}\) to the signal is at least of second order in derivatives of \(M_{ij}\).  Here, the integration constants (in time) \(K_{0i}(\vx)\) and \(K_{00}(\vx)\) are in principle non-zero, and there may be also a constant in time part of \(M_{ij}\), say \(K_{ij}(\vx)\).

At zeroth order in \(\al\), the \(K_{\mu\nu}(\vx)\) components satisfy the massive Poisson's equation without a source term, and are thus determined by the boundary conditions.  Note that in a spherically symmetric system \(K_{0i}(\vx)=0\) by symmetry, and only the trace of \(K_{ij}(\vx)\), \(K(\vx)=K^i_{i}(\vx)=-K_{00}(\vx)\), can be non-zero.  At linear order in \(\al\), there is a source to be added to the \(\Mmn\) wave-equation (\ref{Fieldeq}), that is given by the direct coupling \Eq{eq:action}.  Assuming we can treat \(\al\) perturbatively (as is the case because \(\al\ll1\)), the solution will be linear in \(\al\), so their contribution to the time residuals will be of order \(\al^2\).  Here we neglect such contribution and we work at the leading, linear order in \(\al\).  For the homogeneous background studied  in~\cite{Marzola:2017}, one may set \(K_{00}(\vx)=-K(\vx)=0\) and expect that at shorter scales, where DM structures such as halos form, the components \(K_{\mu\nu}(\vx)\) may have inhomogeneities characterised by a scale that depends on the halo and the environment (as for instance, scales measuring departures from spherical symmetry).  In any case, to estimate their contribution, further study and modeling is clearly necessary.  Nevertheless, even though the \(K_{\mu\nu}\) are constant in time, note the integration goes along the photon path, and they are expected to change over that distance.  On the other hand, a constant in time contribution from \(K_{ij}\) to the frequency shift cannot be measured with PTAs.  Neglecting the contribution of the \(K_{\mu\nu}\) and assuming gradients are \(\pd_i\sim1/\ldB\), since \(\ldB m\gg1\), the integrand in \Eq{eq:freq0_gral2} is expected to oscillate fast and hence to give a small contribution compared to the first two terms.

Let us now add the pulsar term to the result in \Eq{eq:freq}.  Defining the emission time \(t_0 = t-D\) for a pulsar at distance \(D\), we obtain
\begin{align}
	\frac{\nu-\nu_0}{\nu_0}
	& \simeq \frac{\al}{\sqrt2 m\mpl} \left[ \sqrt{\rhoDM}_{\oplus}\vep_{ij,\oplus}\cos\left(mt+\Upsilon_\oplus\right) -\sqrt{\rhoDM}_{\ostar} \vep_{ij,\ostar}\cos\left(mt_0+\Upsilon_\ostar\right) \right] n^in^j \,,
\end{align}
or, after subtracting the average time residual over the observation time,
\begin{align}\label{eq:shift2}
	\trt & = -\frac{\al}{\sqrt2 m^2\mpl} \left[\sqrt{\rhoDM}_{\oplus} \vep_{ij,\oplus}\sin\left(mt+\Upsilon_\oplus\right) - \sqrt{\rhoDM}_{\ostar}\vep_{ij,\ostar}\sin\left(mt-mD+\Upsilon_\ostar\right) \right] n^in^j \,.
\end{align}
For concreteness in what follows we assume that \(\rhoDM\) is given by its average value everywhere and retain the spatial dependence in the polarisation and the phase.  We can consider two possibilities.  We can either assume that the configuration of the ULDM quadrupole is also the same everywhere, beyond its natural dB coherence scale (as implicitly done in, e.g.,~\cite{Nomura:2019cvc} for a spin-1 ULDM model), in which case \(\vep_{ij,\oplus}=\vep_{ij,\ostar}\).  Alternatively, if the quadrupole is different at each pulsar location, we can average over all possible quadrupole configurations, assuming all pulsars live in uncorrelated patches, to obtain the ``helicity average''
\begin{align}\label{eq:shift_full}
	\trHel & = \frac{\al^2 \rhoDM}{2m^4\mpl^2} \left[ \left(\vep_{ij}^\oplus n^in^j\right)^2 \sin^2\left(mt+\Upsilon_\oplus\right) - \frac{2}{15} \sin^2\left(mt-mD+\Upsilon_\ostar\right) \right] \,.
\end{align}
As it should, the pulsar term does not depend on the direction \(\vn\); also, there is no cross-term between Earth and pulsar as it (correctly) averages to zero.  Once again, the phase \(\Upsilon\) in principle is also different for each pulsar: if the characteristic scale for changes of the phase is the same as that of the quadrupole configuration, to account for this effect it would be  appropriate to also average over it.  However, to first order, this is equivalent to averaging over time, and the resulting time-independent contribution is not measurable.  Therefore, keeping the phase here corresponds to a particular assumption where only changes in the quadrupole configuration are significant for the corresponding set of pulsars.

The overall amplitude of the effect, once we average over the sphere \(\vn\), is the same in both cases:
\begin{align}\label{eq:shift_full_avg}
	\trSq &= \frac{\al\sqrt{2\rhoDM}}{\sqrt{15}m^2\mpl} \cos\left(mt-\frac{mD}{2}+\frac{\Upsilon_\oplus+\Upsilon_\ostar}{2}\right) \,.
\end{align}
Thus, the overall effect is at most stronger by a factor of \(\sqrt2\), because there can be a further suppression caused by the different phase for each pulsar.  Therefore, the Earth term dominates at least by a factor of \(1/\sqrt2\), and the effect including both terms is \emph{larger} than if we keep only the Earth term.  Hence, the limits we have obtained above are conservative.

\bibliographystyle{hieeetr}
\bibliography{biblio.bib}
\end{document}